\documentclass[useAMS,usenatbib]{mn2e}

\usepackage{amsmath}
\usepackage{graphicx}

\newcommand{\bfb}{\mbox{\boldmath$b$}}
\newcommand{\bfe}{\mbox{\boldmath$e$}}

\newcommand{\bfu}{\mbox{\boldmath$u$}}

\newcommand{\bfB}{\mbox{\boldmath$B$}}

\newcommand\calE{{\cal E}}

\newcommand{\bfcalE}{\boldsymbol{{\cal E}}}

\newcommand{\bfxhat}{\mbox{\boldmath $\hat x$}}
\newcommand{\bfyhat}{\mbox{\boldmath $\hat y$}}

\newcommand{\ahat}{\mbox{$\hat a$}}

\title[The $\alpha$-effect in rotating convection]{The $\alpha$-effect in rotating convection: a comparison of numerical simulations}
\author[D.W. Hughes, M.R.E. Proctor and F. Cattaneo]{D.W. Hughes$^{1}$\thanks{E-mail:
d.w.hughes@leeds.ac.uk; mrep@cam.ac.uk; cattaneo@flash.uchicago.edu}, M.R.E. Proctor$^{2}$\footnotemark[1]
and F. Cattaneo$^{3}$\footnotemark[1]\\
$^{1}$Department of Applied Mathematics, University of Leeds, Leeds LS2 9JT\\
$^{2}$DAMTP, Centre for Mathematical Sciences, University of Cambridge, Cambridge CB3 0WA\\
$^{3}$Department of Astronomy and Astrophysics and the Computation Institute, University of Chicago, Chicago, IL 60637, USA}

\begin{document}

\date{\today}

\pagerange{\pageref{firstpage}--\pageref{lastpage}} \pubyear{2010}

\maketitle

\label{firstpage}

\begin{abstract}
Numerical simulations are an important tool in furthering our understanding of turbulent dynamo action, a process that occurs in a vast range of astrophysical bodies. It is important in all computational work that comparisons are made between different codes and, if non-trivial differences arise, that these are explained. \citet{KKB10} describe an attempt to reproduce the results of \citet{HP09} and, by employing a different methodology, they arrive at very different conclusions concerning the mean electromotive force and the generation of large-scale fields. Here we describe why the simulations of \citet{KKB10} are simply not suitable for a meaningful comparison, since they solve different equations, at different parameter values and with different boundary conditions. Furthermore we describe why the interpretation of \citet{KKB10} of the calculation of the $\alpha$-effect is inappropriate and argue that the generation of large-scale magnetic fields by turbulent convection remains a problematic issue.
\end{abstract}


\section{Introduction}

One of the most important problems in astrophysical magnetohydrodynamics (MHD) concerns the generation of large-scale magnetic fields, as manifested, for example, by the strong toroidal field of the Sun. This has been most frequently addressed within the theoretical framework of mean field MHD, which addresses the evolution of a large-scale magnetic field through the parameterisation of small-scale turbulent interactions --- leading, for example, to the famous $\alpha$-effect of mean field dynamo theory. With ever-increasing computational power, it has now become possible to explore directly, via numerical simulations, the issue of the generation of large-scale magnetic fields by small-scale turbulence, and hence to compare the results with the predictions of mean field theory. In particular, numerical simulations should help us to answer the question of whether small-scale helical turbulence can lead to the generation of large-scale magnetic fields at high values of the magnetic Reynolds number, $Rm$, the regime of astrophysical relevance. However, far from a consensus emerging from the various computations that have been performed, different groups have come to radically different conclusions. Specifically, certain computations suggest that a large-scale field is indeed generated, whereas others find only small-scale dynamo action, with a marked absence of a large-scale component. If numerical simulations are therefore to be of value, it is important to look in some detail as to why different computations, tackling ostensibly the same problem, come to such dramatically different conclusions. It is clear that different numerical simulations of the same set of equations, with the same parameter values and boundary conditions, should yield the same results --- indeed, to procure such agreement is an important element in the validation of any code. It is doubly important to understand any major disagreement between different studies when they are used to advance contradictory views of the underlying physics. 

A recent example of such a disagreement arises from numerical simulations of convective dynamo action in a rotating plane layer. Specifically, the work of \citet{KKB10}, which claims to solve the same problem as investigated by \citet{HP09}, reaches completely different conclusions concerning the generation of large-scale magnetic fields. The aim of the present paper is to look closely at these two sets of simulations, to point out key differences between them and to discuss the different methodologies employed. 


\section[]{A Case Study}

\subsection[]{Comparing the two models}

In this case study we consider numerical simulations designed to model the evolution of magnetic fields in a turbulent, rotating, convective flow in a plane layer. This particular dynamo problem has been studied for a number of years, from the early pioneering analytical work of \citet{CS72} and \citet{Soward74} to numerical simulations by, for example, \citet{StPierre93}, \citet{JR00}, \citet{RJ02}, \citet{SH04}, \citet{CH06} (paper I hereafter) and \citet{HC08}(paper II). It has been extended by \citet{HP09} (paper III) to incorporate the additional influence of an imposed large-scale shear flow. Recently, \citet{KKB10} (hereafter KKB) have `set out to reproduce the results of \citet{HP09}'; their paper does though include results for the conceptually simpler problem when there is no imposed shear, and thus direct comparison can also be made with the results of papers I and II. It is on this comparison that we shall chiefly focus.

Although the stated aim of KKB was to compare their results with those of \citet{HP09}, there are significant differences in both the models and the methodology employed. Although these differences received little comment in KKB, we believe that it is important to register them from the outset, as they clearly have a bearing on the results themselves and the conclusions subsequently drawn.

Papers I and II considered plane-layer, rotating Boussinesq convection, governed by five parameters: the Rayleigh number $Ra$, the Taylor number $Ta$, the Prandtl number $Pr$, the magnetic Prandtl number $Pm$ and the aspect ratio $\lambda$. In I and II, $Ta$, $Pr$ and $Pm$ took the fixed values $Ta=500\, 000$, $Pr=1$ and $Pm=5$. In I we investigated the cases of $Ra=10^6$, $Ra=5 \times 10^5$ and $Ra=150\, 000$, considering the two aspect ratios of $\lambda=5$ and $\lambda=10$. In II, the main aim was to determine the role of the size of the domain, for various $Ra$, so we considered values of $\lambda$ between $0.5$ and $5$, for three values of the Rayleigh number, $Ra=80\, 000$, $150\, 000$ and $500\, 000$. Papers I and II considered impermeable, stress-free horizontal boundaries that were perfectly conducting, both thermally and electrically; all variables were assumed to be periodic in the two horizontal directions. Paper III started with a case considered in both I and II ($Ra=150\, 000$, $Ta=500\, 000$, $Pr=1$, $Pm=5$, $\lambda=0.5$, with the same boundary conditions) and investigated the influence of an imposed large-scale shear flow.

KKB also considered plane-layer, rotating convection in a domain with $\lambda=5$, though rather than employ the Boussinesq approximation they solved the compressible MHD equations with a weak stratification. They chose the fixed values of $Ra=100\, 000$, $Pr=0.6$ and $Pm=2$, none of which correspond to those used in III. KKB do not specify an input parameter for the rotation rate, as would be necessary for a true comparison is to be made, but instead just give values of the Rossby number and the magnetic Reynolds number, both of which can be determined only \textit{a posteriori} and, even then, only approximately; these suggest that $Ta \approx 6.68\times 10^5$, giving a critical value of the Rayleigh number for the onset of convection as $Ra_c \approx 71\,000$ \citep{Chandra61}. Finally, the boundary condition on the magnetic field on the top and bottom of the domain, namely that the field be prescribed to be vertical on these boundaries, is different from that used in I -- III.

Putting aside for the moment the complicating role of an imposed shear, the results of KKB ought, if any comparison is to be meaningful, to correspond to those of paper III with no shear, a case studied in detail in I and II. This however seems not to be the case, though KKB make little comment on this issue. For the parameter values of I and II, convection sets in when $Ra=59\,008$, but dynamo action does not occur until $Ra \approx 170\, 000$. In this range of $Ra$ it is therefore possible to explore unambiguously the mean electromotive force (emf) due to an imposed mean magnetic field. In the geometry considered in I and II, spatial averages are taken over horizontal planes, leading to a mean emf that is horizontal and depth-dependent. The mean emf in I and II was calculated as the response to an imposed uniform horizontal field. Two of the main findings of I and II were (i) that the emf is highly fluctuating in time, and requires very long temporal averages to pin down its mean values, and (ii) that the mean value is extremely small compared with the size of the fluctuations (of the order of $u_{rms}/Rm$). KKB, on the other hand, reach rather different conclusions, and it is therefore important to understand the possible reasons for their conclusions. We believe that there are three specific issues to explore.

\subsection[]{The parameter regimes considered}\label{subsec:params}

The key result of paper~II was that the form of the convection has a marked effect on the nature of the emf. In particular, ordered motions, as brought about either by convection that is only moderately supercritical, or, alternatively, via constraints imposed by a small aspect ratio, lead to an average emf of much greater magnitude than do fully turbulent motions, for which the emf is spatially and temporally incoherent and hence has a very small average value. Specifically, in II it was demonstrated that for $Ra = 80\,000$ (i.e.\ $Ra/Ra_c \approx 1.36$), the more ordered, less turbulent convection leads to a greater emf than for the case of $Ra = 150\,000$ (i.e.\ $Ra/Ra_c \approx 2.54$). For the simulations of KKB, $Ra/Ra_c \approx 1.41$, a ratio considerably lower than for the $Ra = 150\,000$ case of II with which they choose to make comparison. Thus, simply from consideration of the vigour of the convection, it is of no surprise that the results are different. It would not be a great surprise either, though we have not explored this, if employing different values of $Pr$ and $Pm$ (as in KKB) also lead to differences in the resulting emf.

\subsection[]{The role of the boundary conditions}\label{subsec:bc}

An important point to note straightaway is that KKB calculate the mean emf in the same manner as in I and II, by imposing a uniform horizontal magnetic field. However, whereas this choice is consistent with the boundary conditions in I and II, even in the absence of fluid motion, it is inconsistent with the boundary conditions in KKB, which stipulate that the field be vertical on the upper and lower boundaries. It would appear, though it is not made clear explicitly, that the stated boundary condition in KKB is enforced on the perturbed field, but not on the mean field; interpretation of the results in terms of large-scale field generation would though then seem problematical.

Figure~5a of KKB shows a contour plot of the $\alpha$-effect with depth; both the spatial distribution and the magnitude of $\alpha$ deserve comment. The $\alpha$-effect is antisymmetric about the mid-plane, which one would expect for Boussinesq convection; it is though maximal on the upper and lower boundaries, which one would certainly not expect. For impermeable, perfectly electrically conducting boundaries, as employed in I -- III, it is straightforward to show that the horizontal components of the emf must vanish on the upper and lower boundaries; this is in keeping with considerations of the reflectional symmetry of the convection --- the kinetic helicity, for example, vanishes on the upper and lower boundaries. For the convective flows of KKB, it remains the case that the kinetic helicity vanishes on the upper and lower boundaries. It is therefore surprising that the horizontal components of the emf do not, but this is presumably due to the inconsistency inherent in the application of the magnetic field boundary condition. It is also noteworthy that whereas the $\alpha$ effect in I--II was very small ($O(u_{rms}/Rm)$), in KKB it is large ($O(u_{rms})$), though confined to narrow boundary layers.

The significance of the choice of boundary conditions is readily demonstrated by the following example of the calculation of the emf in a simple kinematic flow. Consider a two-dimensional incompressible flow between the planes $z=0$ and $z=1$, with velocity given by
\begin{equation}
\bfu = \nabla \times \psi \bfyhat, \qquad \textrm{with} \quad
\psi = \sin kx \sin \pi z .
\end{equation}
Suppose a uniform magnetic field of strength $B_0$ is imposed in the $x$-direction, and, for simplicity, that it is sufficiently weak that it can be treated kinematically. Furthermore, suppose that the magnetic Reynolds number is small, so that products of fluctuations can be ignored in the induction equation (first order smoothing). The (steady) fluctuation magnetic field $\bfb$ then satisfies
\begin{equation}
\bfB_0 \cdot \nabla \bfu + \eta \nabla^2 \bfb = 0 .
\label{eq:fluct_ind}
\end{equation}
Writing $(b_x(x,z),0,b_z(x,z)) = \nabla \times a(x,z) \bfyhat$, with $a(x,z)= \ahat(z) \cos kx$, equation~(\ref{eq:fluct_ind}) becomes
\begin{equation}
\left( \frac{d^2}{dz^2} - k^2 \right) \ahat = - \, \frac{B_0 k}{\eta} \sin \pi z,
\label{eq:a_ind}
\end{equation}
with general solution
\begin{equation}
\ahat = \alpha \sinh kz + \beta \cosh kz + \frac{B_0 k}{\eta (k^2+\pi^2)} \sin \pi z.
\label{eq:a_soln}
\end{equation}
Perfectly conducting horizontal boundaries (as assumed in I--III) dictate that $\ahat = 0$ at $z=0$ and $z=1$, with solution
\begin{equation}
\ahat(z) = \frac{B_0 k}{\eta (k^2+ \pi^2 )} \sin \pi z .
\label{eq:bc_pc}
\end{equation}
Boundary conditions for which the fluctuating field $\bfb$ is vertical on the horizontal boundaries (which is, we believe, the boundary condition in KKB) require that $d \ahat /dz = 0$ at $z=0$ and $z=1$, with solution
\begin{align}
\ahat(z) = \frac{B_0}{\eta (k^2+ \pi^2 )}
\Biggl( \pi &\left( \frac{1+\cosh k}{\sinh k} \right) \cosh kz \nonumber \\
&- \pi \sinh kz + k \sin \pi z \Biggr) .
\label{eq:bc_mc}
\end{align}
The mean emf $\bfcalE = \langle \bfu \times \bfb \rangle$ may then be calculated for these two different choices of boundary conditions. To relate the results to those in I--III and in KKB, we define the average as being over $x$, thus giving a $z$-dependent mean emf. For perfectly conducting boundaries, $\calE$ ($=\calE_y$, the only non-zero component) is given by
\begin{equation}
\calE(z) = - \, \left( \frac{B_0 \pi k^2}{2 \eta (k^2+\pi^2)} \right) \sin 2 \pi z ,
\label{eq:emf_pc}
\end{equation}
whereas for the other choice of boundary condition,
\begin{align}
\nonumber
\calE(z) = &- \, \frac{B_0 k}{2 \eta (k^2+\pi^2)} \Biggl( 
\pi \left( k \sin \pi z \cosh kz + \pi \cos \pi z \sinh kz \right) \\
\nonumber
&- \pi \left( \frac{1+\cosh k}{\sinh k} \right) \left( k \sin \pi z \sinh kz + \pi \cos \pi z \cosh kz \right) \\
&- k \pi \sin 2 \pi z \Biggr) .
\label{eq:emf_mc}
\end{align}
\begin{figure}
\begin{center}
\includegraphics[bb = 120 340 596 780, width = 7.5cm]{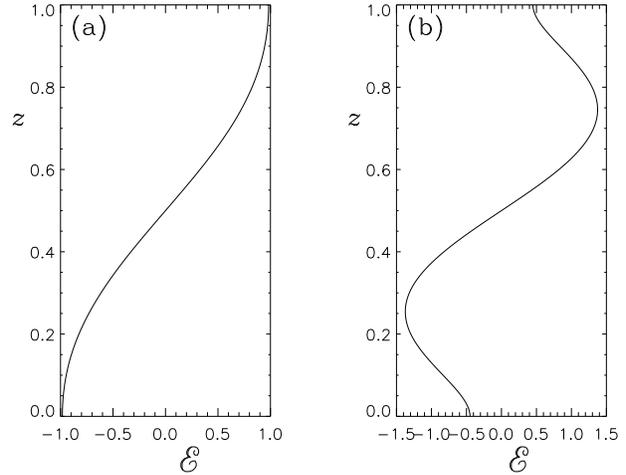}
\end{center}
\caption{emf versus $z$ from expression (\ref{eq:emf_mc}) (i.e.\ for the KKB boundary condition). (a) $k=1$; (b) $k=10$.}
\label{fig:emf_mc}
\end{figure}
Clearly the $z$-dependence of the two emfs (\ref{eq:emf_pc}) and (\ref{eq:emf_mc}) is quite different and that, whereas the emf (\ref{eq:emf_pc}) vanishes at horizontal boundaries that are perfectly conducting, the emf (\ref{eq:emf_mc}) (with a vertical magnetic perturbation at the boundaries) does not. Indeed, as can be seen from Figure~\ref{fig:emf_mc}, the emf can be maximised in strength at the boundaries, reminiscent of the result of KKB (though for our simple model this depends on the value of $k$). The choice of boundary conditions is thus of crucial importance in determining the form of the emf; it is therefore not surprising, on consideration of the boundary conditions alone, that the results of KKB differed from those of I--III.

\subsection[]{Calculating the $\alpha$-effect}\label{subsec:alpha}
\begin{figure}
\begin{center}
\includegraphics[bb = 99 389 596 1100, width = 8cm]{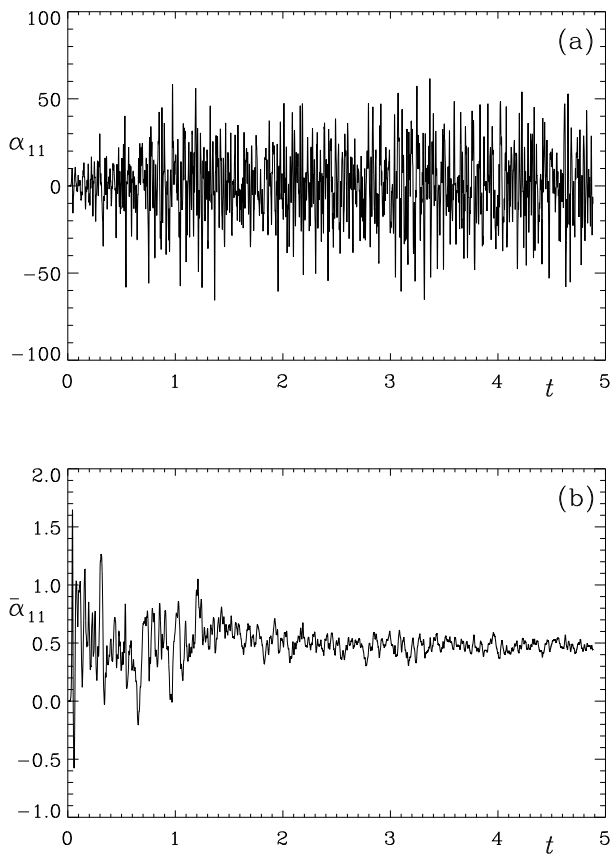}
\end{center}
\caption{(a) $\alpha_{11}$ and (b) its cumulative average $\bar \alpha_{11}$, versus time for plane layer, turbulent, rotating, Boussinesq convection; $Ra=150,000$, $Ta=500,000$, $Pr=1$, $Pm=5$, $\lambda=5$. The time scale is the thermal diffusion time across the layer.}
\label{fig:alpha}
\end{figure}

\begin{figure}
\begin{center}
\includegraphics[bb = 180 382 596 800, width = 6cm]{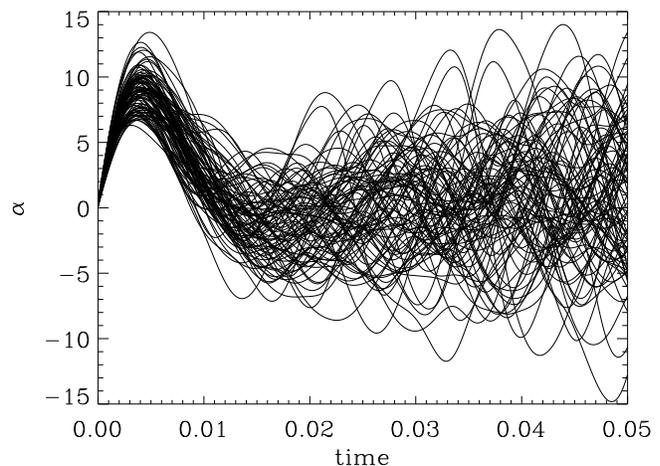}
\end{center}
\caption{Ninety realisations of the kinematic $\alpha$-effect for turbulent rotating Boussinesq convection with a `resetting time' of $0.5$; $Ra=150,000$, $Ta=500,000$, $Pr=1$, $Pm=5$, $\lambda=5$.}
\label{fig:alpha_reset}
\end{figure}
The third issue to discuss is somewhat different, in that it concerns not the set-up of the model (as in \S\S \ref{subsec:params}, \ref{subsec:bc}), but rather the means of determining the $\alpha$-effect and, by implication, its physical interpretation. In I--III and in KKB, a uniform horizontal magnetic field $B_0 \bfxhat$ is imposed and the resulting electromotive force measured. Horizontal averages lead to a $z$-dependent mean emf, which is antisymmetric for Boussinesq convection. The tensor $\alpha_{ij}$ is then defined via the relation
\begin{equation}
\calE_i = \alpha_{ij} B_{0j} ,
\label{eq:alpha}
\end{equation}
where $\bfcalE$ is obtained from half-layer averages (full layer averages vanishing by the Boussinesq symmetry); in equation~(\ref{eq:alpha}), $\calE_i$ (and hence $\alpha_{ij}$) is a function of time. Figure~\ref{fig:alpha}, which reproduces one of the main results from \citet{CH06}, plots $\alpha_{11}$ and its cumulative temporal average versus time, following the imposition of an extremely weak (kinematic) mean field. For the parameter values chosen there is no dynamo action, and hence the mean emf results entirely from the imposed field. Note from Figure~\ref{fig:alpha}(a) that $\alpha_{11}$ fluctuates strongly in time, even though at any instant it already involves an average over many convective cells. This is a reflection of the fact that, even though the flow has significant helicity, there is little coherence in the emfs produced locally from the turbulent convective cells, an idea pursued further by \citet{CHT09}. This has two important consequences, illustrated by Figure~\ref{fig:alpha}(b): one is that determining a meaningful value for $\alpha_{ij}$ requires both large spatial and long temporal averages; the other is that the values of $\alpha_{ij}$ that emerge are small, namely $O(u_{rms}/Rm)$, and not $O(u_{rms})$ as might be expected from a simple turbulent scaling argument (for the parameter values of Figure~\ref{fig:alpha}, $u_{rms} \approx 56$ and $Rm \approx 280$.) The temporal averaging is to be regarded as a proxy for spatial averaging; in theory, though certainly not in practice, as discussed in II and further in \cite{CH09}, in a large enough domain the $\alpha$-effect can be pinned down solely from spatial averaging. 
KKB however reject this standard and clearly meaningful method of establishing the correlation between the emf and the mean field --- and, presumably, do not attach any physical significance to the resulting value of $\alpha$. Instead, they argue that it is necessary to perform what they describe as `resetting'. The precise formulation is not given in detail but involves, we believe, averaging $\alpha_{ij}$ over some arbitrarily chosen time interval $\Delta T$, and then starting again with a new imposed field $\bfB_0$; their desired value of $\alpha_{ij}$ then results from a further averaging of all of these `reset' $\alpha_{ij}$.

In order to clarify this idea, we have calculated $\alpha_{ij}$ for the Boussinesq system of I-II, with various resetting times. As an example, Figure~\ref{fig:alpha_reset} shows $\alpha_{11}$ versus time for ninety consecutive realisations with a resetting time of $\Delta T = 0.05$ (this is slightly longer than an advective timescale across the layer, $t_{ad} \approx 0.02$). From inspection of Figure~\ref{fig:alpha_reset} it is clear that a very short resetting time (less than $t_{ad}$) would lead to a sizeable value of $\alpha$; for the parameters of Figure~\ref{fig:alpha_reset}, $\Delta T = 0.005$ leads to $\alpha_{11} \approx 6.7$, averaged over all the trajectories. For $\Delta T = 0.05$, $\alpha_{11}$ falls to $\alpha_{11} \approx 1.3$; it will continue to fall as $\Delta T$ is increased until it reaches its long-time average of $0.5$ shown in Figure~\ref{fig:alpha}(b). These results are not particularly surprising. Taking $\Delta T < t_{ad}$ simply captures the initial transient behaviour of $\alpha$, as described by the short sudden approximation; this reflects the helicity of the flow but has no knowledge of diffusion. KKB claim that obtaining a value of $\alpha$ that is independent of $Rm$ is `in accordance with mean field theory', although mean field theory, in general, certainly does not neglect diffusion. Over times longer than an advective time, $\alpha$ is wildly fluctuating in time. Thus increasing $\Delta T$ decreases the calculated values of $\alpha$, and it is possible to obtain any value between that when $\Delta T \approx t_{ad}$ (here $\alpha_{11} \approx 6.67$) and that when $\Delta T$ is of the order of the Ohmic time (i.e.\ here $\Delta T = 5$ leading to $\alpha_{11} \approx 0.5$).

The crucial question then is what meaning, if any, should be ascribed to this range of possible values. We maintain that it is only the long-time value that accounts for all aspects of the $\alpha$-effect, including diffusion, and that it is this value that has its traditional physical meaning of determining the growth (or decay as in this case) of a long wavelength magnetic field perturbation. There seems to be no justification for taking any of the larger values of $\alpha$ that might emerge via a `resetting' procedure. Indeed, for the example that we are considering, it is easy to see that a value of $\alpha$ that is independent of $Rm$ cannot be correct. As argued in I, for the plane layer convective system under discussion, the horizontally averaged induction equation can be written as
\begin{equation}
\frac{\partial}{\partial t} \langle {\bfB} \rangle = {\bfe}_z \times \frac{\partial}{\partial z} {\bfcalE} 
+ \eta \frac{\partial^2}{\partial z^2}\langle {\bfB} \rangle ,
\label{h_average}
\end{equation}
where it is important to notice that the diffusion coefficient is simply $\eta$ and not $\eta + \beta$, with $\beta$ the eddy diffusivity \citep{CS72}; this follows from the anisotropic nature of the averages arising from the existence of a separation of scales in the horizontal but not in the vertical. Now for the parameter values of Figure~\ref{fig:alpha} there is an $\alpha$-effect, but no large-scale dynamo action (indeed, no dynamo of any kind). It follows immediately therefore that any $\alpha$-effect cannot exceed a number of order $\eta$, which is entirely consistent with the result of Figure~\ref{fig:alpha}(b) and the conclusions of I--II, but inconsistent with the claims of KKB.

\section[]{Discussion}

The stated aim of KKB was to `reproduce the results of Hughes \& Proctor (2009)' (and hence, for the case of no shear, those of I and II). For reasons clearly explained in \S\ref{subsec:params} and \S\ref{subsec:bc}, this plan was doomed from the outset: the choices of parameter values and boundary conditions are both important and need to be treated with respect. That is not to say that it is not of interest to explore the consequences of changing the parameter values and the boundary conditions --- it certainly is --- but that was not the thrust of KKB. Furthermore, it is our view that the KKB calculation of the emf through the imposition of a magnetic field that does not satisfy the stated boundary condition is mathematically inconsistent. Calculating the $\alpha$-effect is not an end in itself; it is useful only if it can provide information about large-scale field evolution. But if the emf is derived from the imposition of a magnetic field that does not satisfy the boundary conditions then it is doubtful if the ensuing $\alpha$ has a meaningful physical interpretation.

With the so-called resetting procedure to measure the $\alpha$-effect it is indeed true that one can obtain a more significant value of $\alpha$. But is this greater $\alpha$-effect meaningful? For the convective dynamos discussed here, the answer is clearly no, as explained above. Indeed, if temporal averaging is a proxy for inadequate spatial averaging, then the very idea of a resetting time has no meaning.

\section*{Acknowledgments}

This research was supported by the Science and Technology Facilities Council and by the National Science Foundation sponsored Center for Magnetic Self Organization (CMSO) at the University of Chicago.

\bsp

\label{lastpage}

\end{document}